\documentclass[twocolumn,showpacs,amsmath,amssymb]{revtex4}
\usepackage{graphicx}
\usepackage{dcolumn}
\usepackage{bm}
\usepackage{amsmath}
\usepackage{amsfonts}
\begin{document}

\title{Growth of Black Holes in the interior of Rotating Neutron Stars}
\author{Chris {\sc Kouvaris}}\email{kouvaris@cp3.sdu.dk}
\affiliation{$\text{CP}^3$-Origins, University of Southern Denmark, Campusvej 55, Odense 5230, Denmark}
\author{Peter {\sc Tinyakov}}\email{Petr.Tiniakov@ulb.ac.be}
 \affiliation{Service de Physique Th\'eorique,  Universit\'e Libre de Bruxelles, 1050 Brussels, Belgium}
 
\begin{abstract}
Mini-black holes made  of dark matter that can potentially form in the
interior of neutron stars have been always thought to grow by accreting the
matter of the core of the star via a spherical Bondi accretion. However,
neutron stars have sometimes significant angulal velocities that can in
principle stall the spherical accretion and potentially change the conclusions
derived about the time it takes for black holes to destroy a star. We study
the effect of the star rotation on the growth of such black holes and the
evolution of the black hole spin. Assuming no mechanisms of angular momentum
evacuation, we find that even moderate rotation rates can in fact destroy
spherical accretion at the early stages of the black hole growth. However, we
demonstrate that the viscosity of nuclear matter can alleviate the effect of
rotation, making it possible for the black hole to maintain spherical
accretion while impeding the black hole from becoming maximally
rotating.  \\[.1cm] {\footnotesize \it Preprint: CP$^3$-Origins-2013-049
  DNRF90 \& DIAS-2013-49.}
\end{abstract}


\maketitle 

\section{Introduction}
\label{sec:introduction}

A lot of effort has been devoted by theoretical and experimental
physicists in order to unveil the mystery of dark matter. One possible
way to indirectly observe the existence or constrain the properties of
dark matter is by looking on possible effects dark matter might have
on stars. This includes for example constraints on dark matter through
asteroseismology
~\cite{Lopes:2012af,Casanellas:2012jp,Casanellas:2013nra},
modification of the transport properties of the
star~\cite{Frandsen:2010yj,Horowitz:2012jd}, exotic new
effects~\cite{PerezGarcia:2010ap,PerezGarcia:2011hh}, and hybrid dark
matter rich compact stars~\cite{Leung:2013pra,Goldman:2013qla}. For
compact stars such as neutron stars and white dwarfs there are
additional ways to constrain dark matter either by studying the effect
of Weakly Interacting Massive Particles (WIMP) annihilation in the
core of the star on the star's
temperature~\cite{Kouvaris:2007ay,Bertone:2007ae,Kouvaris:2010vv,deLavallaz:2010wp}
or --- in the case of asymmetric dark matter --- by studying the
possibility of forming a mini-black hole out of WIMPs that eventually
destroys the host compact star. This is the possibility that we
entertain in this paper.

Under certain circumstances, the compact star might accumulate a
sufficient number of WIMPs which, after losing their initial kinetic
energy, thermalize with the surrounding nuclear matter in the core of
the star. Depending on the nature of the particle (i.e., boson or
fermion), the existence or not of self-interactions, and the mass, the
WIMPs might become self-gravitating and collapse forming a black hole
in the center of the star. Once the black hole is formed, its fate is
determined by the relative strength of two competing processes: the
accretion of matter from the surrounding nuclear matter onto the black
hole, and the Hawking evaporation. The initial mass of the black hole
dictates the result of the competition. Since the accretion rate
increases with increasing black hole mass while the Hawking radiation
decreases, the black holes that are lighter than a certain critical mass
evaporate, while those which are heavier grow. 

The outcome in the two cases is completely different. The evaporation
of a black hole can possibly heat up the star slightly, which is
difficult to observe, if possible at all. On the contrary, the growth
of the black hole leads to the eventual star destruction. Based on
this catastrophic scenario first studied in~\cite{Goldman:1989nd},
severe constraints have been imposed on the mass and WIMP-nucleon
cross section for the bosonic asymmetric dark
matter~\cite{Kouvaris:2011fi,McDermott:2011jp,Guver:2012ba,Bell:2013xk,Bramante:2013hn,Bramante:2013nma},
and for asymmetric fermionic dark matter with attractive
interactions~\cite{Kouvaris:2011gb}. Based on similar scenarios,
constraints on primordial black holes as dark matter have been imposed
in~\cite{Capela:2012jz,Capela:2013yf} as well as constraints on the
spin-dependent WIMP-nucleon cross section~\cite{Kouvaris:2010jy} (in
the latter case,  white dwarfs have been considered instead of
neutron stars).

However, in order for these constraints to be valid, one must make sure
that every single step in this chain of events --- the WIMP capture,
thermalization, collapse, and growth of resulting black hole --- takes
place. Failure in any of these steps might lead to the complete
invalidation of the constraints. In view of this, several issues in
the thermalization that can potentially affect the constraints on the
light WIMPs were examined in~\cite{Bertoni:2013bsa}, as well as some
issues regarding the number of modes emitted in  Hawking
radiation~\cite{Fan:2012qy}. In addition, it has been pointed out
that for non-self-interacting asymmetric bosonic WIMPs, no constraints
can be imposed on heavy WIMPs with masses in the TeV range or higher because the
collapse of the WIMPs in this case would lead to the creation of successive small
black holes that evaporate instead of a single large one that would
grow~\cite{Kouvaris:2012dz,Kouvaris:2013awa}. Finally, similar arguments have been used
in the safety assessment report for LHC~\cite{Giddings:2008gr} in the
context of the hypothetical possibility of  black hole formation in
particle collisions at the LHC.

There is a common element in all the constraints that have been
derived so far. It has been assumed that once the black hole forms,
the accretion of matter onto the black hole proceeds via the so-called
Bondi accretion which gives a large accretion rate scaling as $M^2$
with the black hole mass. However, the Bondi solution is based upon
the assumption of spherical accretion. This means that matter falls
into the black hole isotropically. If the falling matter carries 
angular momentum, it may change the picture completely. It is a well
known fact in astrophysics that if the infalling matter possesses angular
momentum, it forms a disc around the black hole rather
than falling in isotropically. This can significantly
diminish the accretion rate and therefore could change or even
invalidate the derived constraints in two ways: the Hawking radiation might
dominate for heavier black holes, and/or the accretion rate may be too
slow to lead to the star destruction in the star lifetime. 

In reality, the ideal conditions for Bondi accretion are never met
since all compact stars, and in particular neutron stars always
rotate, sometimes with high angular velocities. The rotating
in-falling matter not only can stall the accretion due to angular
momentum, but it can also spin up the black hole to maximum rotation
rates, thus changing the accretion rate as compared to the (non-rotating)
Bondi case. So, the effect of rotation requires a detailed study,
which is the purpose of this paper. In fact, as we will show, in
realistic cases rotation has an important effect, and the conditions
for the Bondi accretion are not met automatically. We will demonstrate
that only if one considers the effect of the viscosity of nuclear
matter in the core of the neutron star, one can recover the conditions
for Bondi accretion and thus save the imposed constraints.

\section{The effect of rotation}
\label{sec:effect-rotation}

Two main issues have to be addressed in order to make sure that the
neutron star rotation does not change substantially the estimates for the
accretion rate. First, one has to check whether the Bondi accretion
regime is valid through the entire star consumption process. Second, 
it has to be checked that the black hole itself does not become
maximally rotating and the Schwarzschild solution remains a good
approximation.

\subsection{Validity of the Bondi regime}
\label{sec:valid-bondi-regime}

In the Bondi regime, the accreted matter is characterized by an
$r$-dependent energy density, pressure and velocity~\cite{Shapiro:1983du}. This
description holds down to the Bondi radius $r_s$,
\begin{equation}
r_s = {G M\over 4 c_s^2},
\label{eq:r_s}
\end{equation} 
where $M$ is the black hole mass and $c_s=0.17$ is the sound speed of matter in the core of a neutron star far away
from the black hole. At $r<r_s$ the flow becomes supersonic. Note that
in the case of a black hole inside a neutron star the sound speed is a finite
fraction of the speed of light, and the Bondi radius is only a few times
larger than the black hole horizon size.

Ideally, the Bondi accretion is spherically symmetric and assumes zero
vorticity of matter collapsing into the black hole. However, all stars rotate
to some extend, and in particular neutron stars may have rotation periods as
short as milliseconds. Rotation of matter falling into the black hole can
destroy the conditions for the Bondi accretion. Due to  conservation of
angular momentum an accretion disc can be formed reducing significantly the
accretion rate. This may invalidate the constraints that are based on the star
destruction as an observable effect, because there may  not be enough time for
a black hole to consume the whole star. We show below that, although the
rotation cannot be ignored, the conditions for the Bondi accretion are still
valid.

The rotation cannot break the Bondi accretion regime if the accreted matter
reaches the innermost stable orbit with the angular momentum much smaller than
the keplerian angular momentum it would have at that orbit. The keplerian specific
angular momentum at the innermost stable orbit is
\begin{equation}
l_{\text{iso}}=2\sqrt{3} \psi GM,
\end{equation}
where $\psi$ is 1 for a nonrotating Schwartzscild black hole, and $1/3$ for an
extreme Kerr black hole. Considering the worst case of matter near the
equatorial plane, the specific angular momentum of a piece of matter at a
distance $r_0$ from the center of the star rotating with an angular velocity
$\omega_0$ is $l=\omega_0 r_0^2$.  At the time this piece of matter reaches
the innermost stable orbit, all the matter at smaller radii has been already
accreted, so that the mass of the black hole is $M=4/3 \pi \rho_c r_0^3$,
where we have neglected the initial mass of the black hole (which was simply the mass of the collapsed WIMP population that triggered the formation of the black hole). The condition
$l<l_{\text{iso}}$ translates into the condition $M>M_{\rm crit}$ for the mass
$M$ of the black hole. In other words, once the black hole mass grows beyond
the critical value
\begin{equation}
M_{\text{crit}}=\frac{1}{12^{3/2}} 
\left (\frac{3}{4\pi\rho_c} \right )^2 
\left ( \frac{\omega_0}{G} \right )^3 \frac{1}{\psi^3},
\label{eq:Mcrit}
\end{equation}
the angular momentum of the accreting matter cannot stall the
accretion and can be safely ignored. Using a typical value
$\rho_c= 5 \times 10^{38}~\text{GeV}/\text{cm}^3$, we find 
\begin{equation}
M_{\rm crit}=2.2 \times 10^{46}P_1^{-3}~{\rm GeV},
\label{eq:Mcrit}
\end{equation}
where $P_1$ is the star rotation period $P$ measured in seconds. In
practice, we are interested in constraints from nearby old neutron
stars (e.g. J0437-4715 and J2124-3358) that have periods of $P \sim
5$~ms. For such a period one finds $M_{\text{crit}}=1.7 \times
10^{53}~{\rm GeV} \sim 10^{-4} M_\odot$.

Consider now the growth of the black hole mass from the initial value to
$M_{\rm crit}$. It is clear from the above discussion that rotating infalling
matter can and will stall the accretion as long as there is no mechanism of
getting rid of the extra angular momentum. There are several mechanisms that
can potentially achieve this. We need to demonstrate that at least one can
reduce efficiently the angular momentum of infalling matter. Viscosity can in
fact play this role. It was pointed out in~\cite{Markovic:1994bu}, that in the
case of a black hole accreting matter via Bondi accretion inside a star,
viscosity can enforce an essentially rigid rotation of matter (with
$\omega=\omega_0$) for radii larger than
\begin{equation} 
r_{\nu}= \frac{G^2M^2}{c_s^3\nu}, 
\label{rn}
\end{equation} 
where $\nu$ is the kinematic viscosity of the star's matter and $M$ is
the mass of the black hole. For radii smaller than $r_{\nu}$, the
viscosity cannot brake efficiently the rotation, so that the angular
velocity grows with the decreasing radius as follows from angular momentum
conservation, $\omega= \omega_0 r_{\nu}^2/r^2$. The kinematic
viscosity (assuming neutron superfluidity) for a typical neutron star
is~\cite{riper}
\[
\nu = 2 \times 10^{11}\,T_5^{-2}~\text{cm}^2/\text{s} ,
\]
where $T_5=T/(10^5{\rm K})$, $T$ being the temperature of the star. 

Because of the viscosity, a given piece of nuclear matter falls into
the black hole rotating with a constant angular velocity $\omega_0$ as
long as its distance to the black hole is larger than both $r_\nu$ and
$r_s$. If any of these conditions is violated, we will assume
(conservatively) that the angular momentum is conserved. 

One has to distinguish two cases: $r_{\nu}<r_s$ and the opposite one. As it
can be seen from Eq.~(\ref{rn}), $r_{\nu}$ grows with $M$. By comparing
$r_{\nu}$ and $r_s$ one can verify that the initial black hole (right after
the collapse of the WIMP sphere) falls into the first case for all WIMP masses
in the range between keV to tens of GeV.  The nuclear matter that is consumed
in this regime has specific angular momentum $l=\omega_0r_s^2$. The condition
$l<l_{\text{iso}}$ even for a fast rotating $\sim 5$~ms star is satisfied for
$M<8 \times 10^{57}$~GeV, which is always true.  So, in the first regime the
Bondi accretion is not in danger.  

However, as the mass of the black grows, so
does $r_\nu$. The transition to the regime $r_{\nu}>r_s$ takes place when the
black hole mass becomes $M=M_\nu$ with
\begin{equation}
M_\nu \simeq 2.1 \times 10^{51} T_5^{-2} {\rm GeV}.
\label{eq:Mnu}
\end{equation}
Once the black hole gets heavier than $M_\nu$, the specific angular momentum
of the consumed matter becomes $l=\omega_0 r_{\nu}^2$.  Once again, the Bondi
accretion holds as long as $l<l_{\text{iso}}$.  This leads to the condition
\[
M< M_B 
\]
where 
\begin{equation}
M_B = {c_s^2\over G} \left({2\sqrt{3}\nu^2\over \omega_0}\right)^{1/3} =
2 \times 10^{54} P_1^{1/3}T_5^{-4/3}  {\rm GeV}.
\label{eq:mBondi}
\end{equation}
Here we have set $\psi=1$ assuming a slowly rotating black hole.
For a 5~ms pulsar, this
condition becomes
$M< 3.4 \times 10^{53} T_5^{-4/3}  {\rm GeV}$.
Above this value of the black hole mass the viscosity is not efficient
in evacuating the angular momentum and the Bondi accretion may not be
valid.

To summarize, from the collapse of the WIMP sphere up to the mass
$M_B$ of Eq.~(\ref{eq:mBondi}), the black hole grows in the
Bondi regime because the angular momentum is efficiently evacuated by
the viscosity. On the other hand, for black holes heavier than
$M_{\rm crit}$ of Eq.~(\ref{eq:Mcrit}), the angular momentum of
infalling matter is not sufficient to stall the Bondi accretion. So, the Bondi
regime can only be violated for  black hole masses in the range $M_B<M<
M_{\rm crit}$. 

At temperatures of $T=10^5{\rm K}$ characteristic of old neutron stars, one
actually has $M_{\rm crit} < M_B$, so accretion proceeds always in the Bondi
regime.  But for a rapidly rotating warmer star with the period $P \sim 5$~ms and
the temperature $10^7$K, one has $M_{\rm crit} > M_B$ and the viscosity stops
being effective when the black hole mass satisfies $M_B<M< M_{\rm crit}$. At
this stage of the growth, and for this neutron star temperature the Bondi accretion does
not hold anymore. However note that this is the last and the shortest stage
of the star destruction: were the Bondi regime still valid, the rest of the
star would have been consumed in about a second. So, it is clear that the star
destruction cannot be substantially postponed even if the Bondi regime is not
valid.

Although it is difficult to calculate the exact accretion rate at this
stage, one can easily convince oneself that the star destruction is
eminent and the accretion cannot be stalled for a long time. To make
the argument, recall that so far we have based our estimates on the
worst possible case of the accretion of matter from the equatorial
plane. In Eq.~(\ref{eq:mBondi}) we estimated the black hole mass $M_B$
above which the specific momentum at the equator satisfies $l>l_{\rm
  iso}$. However, matter along the rotation axis carries smaller
specific angular momentum. Keeping in mind that for $r>r_\nu$  
matter rotates with  angular velocity $\omega_0$, the specific
angular momentum of matter at $r=r_\nu$  is $l=\omega_0 r_{\nu}^2 \sin^2\theta$ where $\theta$
is the polar angle i.e. the angle formed between the rotation axis and the line that connects the center of the star with the particular piece of matter in consideration. For the
sake of the argument let us assume that although the Bondi accretion
might have ceased for matter around the equator,  matter at 
higher latitude that satisfies $l<l_{\rm iso}$ can still be
accreted with a Bondi rate.  Under this assumption, the overall
accretion rate is suppressed, as compared to the Bondi case, by the
ratio of the solid angle of the polar cups where the condition 
$l<l_{\rm iso}$ holds to $4\pi$. It is easy to show
that $l<l_{\rm iso}$ as long as $\sin^2\theta < (M_B/M)^3$. The
rate therefore can be written as follows,
\begin{equation}
\frac{dM}{dt}= {4 \pi\lambda \rho_c G^2 \over c_s^3}  
M^2f(M),
\label{eq:bondi-modified}
\end{equation}
where $\lambda$ is a parameter equal to 0.707 for the polytropic neutron star equation
of state with an index $\Gamma=4/3$, and the suppression factor $f(M)$ is 
\begin{equation}
f(M) = 1 - \sqrt{1 - M_B^3/M^3 }.
\label{eq:f-factor}
\end{equation}
The Bondi rate is recovered by setting $f(M)=1$. 

From Eq.~(\ref{eq:bondi-modified}), the growth time $t$ of the black hole from
$M=M_B$ to some $M > M_B$ is given by the equation
\begin{equation}
t = \tau \int_1^{M/M_B} {d\xi \over \xi^2 \left( 1- \sqrt{1-1/\xi^3 }\right) },
\label{eq:growth-time}
\end{equation}
where 
\begin{equation}
\tau = {c_s^3\over 4\pi\lambda \rho_c G^2 M_B} 
\label{eq:tau}
\end{equation} 
is the time scale of the process. For the worst case of 
$P=5$~ms and $T=10^7$~K one has 
$M_B \sim 7.4\times 10^{50}$~GeV and $\tau \simeq 0.7$~s. 
In the Bondi case one has to omit the factor in the brackets in
denominator of Eq.~(\ref{eq:growth-time}). In this case one would have
$t\sim \tau$, as stated above. With the reduced accretion rate, the
growth to some $M\gg M_B$ takes time
\begin{equation}
t \sim \left ( {M\over M_B} \right )^2 \tau. 
\label{eq:growth-time-final}
\end{equation} 
As explained above, given that the black hole has to grow in this
regime only by $2-3$ orders of magnitude until it reaches $M_{\rm
  crit}$, and that $\tau \lesssim 1$~s, the total growth time remains
short (of the order of hours), and the neutron star destruction is eminent.

\subsection{Evolution of the black hole spin}
\label{sec:evolution-black-hole}

Let us now show  that the accreting matter does not make the black hole
maximally rotating. The parameter that characterizes how fast the black hole
rotates is
\[
a=J/J_{\text{crit}},
\]
where $J_{\text{crit}}=GM^2$. Maximally
rotating black hole corresponds to $a=1$.

We first observe that black holes are formed from a BEC state with a
very small angular momentum, $a\ll 1$. Approximating for simplicity the WIMP
sphere in the condensed state as a rigid homogeneous sphere with a moment of
inertia $I=2/5Mr_c^2$, $r_c$ (the radius of the condensed
state) is~\cite{Kouvaris:2011fi}
\begin{equation}
r_c = \left (
\frac{8\pi }{3} G \rho_c m^2\right )^{-1/4}\simeq 1.6 \times 10^{-4} \left
(\frac{\text{GeV}}{m} \right )^{1/2}
\text{cm}. 
\label{condensed_ground}
\end{equation}
The angular momentum is therefore $J=I\omega_0$, where $\omega_0$ is the neutron star
angular velocity. The mass of the black hole at birth is
given by~~\cite{Kouvaris:2011fi} 
\begin{equation}
M=\frac{2}{\pi}\frac{M_{pl}^2}{m} ,
\label{chandra}
\end{equation}
where $m$ is the WIMP mass. Assuming the neutron star has a period of rotation of
$5$~ms one finds $a \simeq 0.035\ll 1$.

In the Bondi regime, the growth of the black hole is governed 
 by the equation~\cite{Page:1976ki} 
\begin{equation}
\frac{dM}{dt}= {4 \pi\lambda \rho_c G^2 \over c_s^3}  
M^2-\frac{f(a)}{M^2}, 
\label{mass_evo}
\end{equation} 
where the first term corresponds to the Bondi accretion rate
(cf. Eq.~(\ref{eq:bondi-modified}) with $f=1$) and the second to the Hawking
radiation. As it was pointed out in \cite{Page:1976ki}, the coefficient of the
Hawking radiation depends not only on the number of different particle species
that the black hole can emit, but also on the black hole rotation: the larger
the value of $a$, the higher is the emission rate.

A remark is in order at this point concerning the establishment of the Bondi
accretion.  If Bondi accretion is not established fast, Hawking radiation
might win over the weak initial accretion and lead to evaporation of the black
hole. This might shift to higher values the critical mass that the black hole
needs in order to survive.  However, this is not an important effect since the
hydrodynamic limit establishes very fast. Making use of
Eqs.(\ref{condensed_ground}) and (\ref{chandra}), the dynamical time scale is
given by
\begin{equation} 
t_{dyn} = \sqrt{\frac{r_c^3}{GM}} \simeq 6 \times 
10^{-10}\,  \text{s} \,\left ( \frac{\text{GeV}}{m} \right )^{1/4}, \label{dyn}
\end{equation}
which is many orders of magnitude smaller than the evaporation times of
black holes that are formed from WIMPs of masses in the keV -- GeV range.  

Consider now the evolution of the black hole spin during the accretion. It is
easy to see that if one ignores  viscosity and assumes  angular momentum
conservation, the black hole can become maximally rotating. Indeed, if a
spherical region of a rotating neutron star of a mass $M$ collapses, the angular
momentum of the resulting black hole scales like $M^{5/3}$, while the
corresponding critical angular momentum is proportional to $M^2$. Therefore,
for sufficiently large $M$ the resulting black hole is subcritical. The
boundary value of the mass $M$ at which an exactly critical black hole would
be formed parametrically coincides with $M_{\rm crit}$ of Eq.~(\ref{eq:Mcrit})
and equals 
\begin{equation}
M_{\rm max} = 5.8\times 10^{46} P_1^{-3}~{\rm GeV}. 
\label{eq:Mmax}
\end{equation}
For the neutron star rotation period of 5~ms
this value equals $4.6\times 10^{53}$~GeV, which is much larger than the
initial mass of the black hole formed from WIMPs. Thus, if the angular
momentum were conserved, the black hole would become critical as soon as it
accretes a mass much larger than the initial one, and would stay close to
critical until the black hole reaches the mass $M\sim M_{\rm max}$. 

The presence of viscosity changes this picture. To see this, consider the
evolution of the parameter $a$. Since $a=J/GM^2$,
the rate of its change is
\begin{equation}
\frac{1}{a}\frac{da}{dt}=\frac{1}{J}
\frac{dJ}{dt}-2\frac{1}{M}\frac{dM}{dt}. 
\label{aa}
\end{equation}
The rate of the change of the black hole mass is given by
Eq.~(\ref{mass_evo}), whereas the rate of change of the momentum is given
by~\cite{Page:1976ki}
\begin{equation}
\frac{dJ}{dt}=F_J -\frac{g(a)J}{G^2M^3}, \label{a_evo}
\end{equation}  
where $F_J$ is the angular momentum accretion rate, and $g(a)$ is a
dimensionless number that characterizes the strength of the
angular momentum loss rate due to  Hawking radiation.

In the presence of viscosity, the matter is accreted with the specific angular
momentum given by either $l=\omega_0r_s^2$ or $l=\omega_0r_{\nu}^2$, whichever
is bigger. As we have seen above, at the first stage when $M<M_\nu$ with
$M_\nu$ given by Eq.~(\ref{eq:Mnu}), one has $r_{\nu}<r_s$. Making use of
$dJ=dM \omega_0 r_s^2$ one finds
\begin{equation}
F_J = \omega_0 r_s^2 {dM \over dt}, 
\label{eq:F_J}
\end{equation}
and thus 
\begin{equation}
\frac{1}{a}\frac{da}{dt}=\frac{1}{J}\omega_0r_s^2\frac{dM}{dt}
-\frac{g(a)}{G^2M^3}-\frac{2}{M}\frac{dM}{dt}. 
\label{j2}
\end{equation}
Together with Eq.~(\ref{mass_evo}) this equation determines the evolution of
the spin parameter $a$ during the accretion until $M=M_\nu$.  
Eq.~(\ref{j2}) can be simplified further. Writing
$J=aGM^2$ and making use of Eq.~(\ref{eq:r_s}) one may cast Eq.~(\ref{j2})
into the following form,
\begin{equation}
\frac{1}{a}\frac{da}{dt} = \left({1\over a} {M\over M_*} -1 \right)
{2\over M} {dM \over dt } -\frac{g(a)}{G^2M^3}, 
\label{eq:da/dt}
\end{equation}
where $M_* = 32c_s^4/(G\omega_0) \simeq 9.6 \times 10^{59} P_1
\text{GeV}$. For $P=5$~ms one has $M_* \simeq 4.8\times 10^{57}{\rm GeV}\simeq
4.8M_\odot$. The last term in Eq.~(\ref{eq:da/dt}) is due to the Hawking
radiation and describes the angular momentum loss. From the previous
discussion we know that this term can only be important at the very first stages
of the black hole growth when the black hole angular momentum is still much
smaller than the critical one. So, this term can be safely ignored.

With the last term omitted, Eq.~(\ref{eq:da/dt}) can be solved analytically in
terms of $a(M)$. Denoting $ M/M_*\equiv \mu$, the solution reads
\begin{equation}
a(\mu) = {\mu_0^2\over \mu^2} a_0 + {2\over 3} {\mu^3 - \mu_0^3\over
  \mu^2}, 
\label{eq:solution1}
\end{equation}
where $a_0\simeq 1.7 \times 10^{-4}/P_1$ and $\mu_0=9.8 \times 10^{-23}/(mP_1)$
are the initial values. The behavior of this solution is quite
simple. Initially, the first term dominates and $a$ decreases. When $\mu$
becomes much larger than $\mu_0$ (i.e. when the black hole mass becomes much
larger than its initial value), the second term eventually starts to dominate
and then
\begin{equation}
a(\mu) \simeq {2\over 3} \mu = {2M \over 3M_*} .
\label{eq:s1-approx}
\end{equation}
This behavior continues as long as Eq.~(\ref{eq:da/dt}) holds, i.e.
until $M=M_\nu$. 

At $M>M_\nu$ Eq.~(\ref{j2}) is modified. The angular momentum accretion rate
now becomes $F_J = \omega_0 r_\nu^2 dM/dt$ with $r_\nu$ given by
Eq.~(\ref{rn}). Thus, the new equation is obtained from Eq.~(\ref{j2}) by the
replacement $r_s\to r_\nu$. It can be rewritten in
a form analogous to Eq.~(\ref{eq:da/dt}),
\begin{equation}
\frac{1}{a}\frac{da}{dt} = \left({1\over a} {M^3\over \tilde M_*^3} -1 \right)
{2\over M} {dM \over dt } -\frac{g(a)}{G^2M^3}, 
\label{eq:da/dt-bis}
\end{equation}
but with a different parameter $\tilde M_* = (2\nu^2/\omega_0)^{1/3} \times
c_s^2/G \simeq 1.6 \times 10^{54} P_1^{1/3}T_5^{-4/3}\text{GeV}$. For
 $P=5$~ms one has $\tilde M_* \simeq 2.8\times 10^{53}T_5^{-4/3}$~GeV. If the last
term is neglected, this equation can also be solved exactly. Denoting again $M/\tilde
M_*\equiv \mu$, the solution is
\begin{equation}
a(\mu) = {\mu_1^2\over \mu^2} a_1 + {2\over 5} {\mu^5 - \mu_1^5\over
  \mu^2}, 
\label{eq:solution2}
\end{equation}
where $a_1$ and $\mu_1$ are the initial values which are obtained by
matching with the solution of Eq.~(\ref{eq:s1-approx}) at the point $M=M_\nu$, 
\begin{eqnarray}
\nonumber
\mu_1 &=& M_\nu /\tilde M_* \simeq 1.2 \times 10^{-3} T_5^{-2/3}P_1^{-1/3},\\
\nonumber
a_1 &=& 2M_\nu /( 3M_*)\simeq 1.45\times 10^{-9} T_5^{-2}P_1^{-1} .
\end{eqnarray}
Note that one has $a_1\ll 1$ for all realistic values of $T_5$ and $P_1$.

The initial values $a_1$ and $\mu_1$ are such that the solution
of Eq.~(\ref{eq:solution2}) grows with $\mu$ and becomes saturated by the term cubic
in $\mu$ already for $\mu$ a few times larger than $\mu_1$. Since we want to
check if the solution can reach $a\sim 1$, this is the regime of interest.
For $\mu\gg \mu_1$ we have at the second stage
\begin{equation}
a(\mu) \simeq {2\over 5} \mu^3 = {2M^3 \over 5\tilde M_*^3} .
\label{eq:s2-approx}
\end{equation}
When the mass reaches the value $M_{\rm max} = 5.8 \times 10^{46}P_1^{-3}$~GeV
given by Eq.~(\ref{eq:Mmax}), beyond which the total angular momentum of
accreted matter is not sufficient to make the black hole maximally rotating,
the angular momentum starts decreasing. Note that, although now we are not
assuming  conservation of  angular momentum, the latter is always
transported from inner to outer layers of the star, and thus black holes with
$M>M_{\rm max}$ are subcritical even in the presence of viscosity.  Thus, the
parameter $a$ can become of order 1 only at $M<M_{\rm max}$. As $M$ grows to
  $M_{\rm max}$, the parameter $a$ grows to its maximum value
\begin{equation}
a_{\rm max} = 2\times 10^{-23} T_5^4/P_1^{10}.
\label{eq:amax}
\end{equation}
As one can see, while generically the parameter $a$ remains small, there is a
strong dependence of its maximum value on the temperature and the period of
the star, and for small periods and large temperatures $a$ may actually reach
the value $a\sim 1$.

It is clear from Eq.~(\ref{eq:s2-approx}) that the problem arises for the
values of masses $M\gtrsim \tilde M_*$. Comparing the definition of $\tilde M_*$ with
$M_B$ of Eq.~(\ref{eq:mBondi}) we see that they are parametrically the same. On the
other hand, the mass at which $a$ reaches the maximum value parametrically
coincides with $M_{\rm crit}$ of Eq.~(\ref{eq:Mcrit}). Therefore, the black hole
spin could reach the values $a\sim 1$ (and thus the accretion rate could be
modified) for the range of masses $M_B\lesssim M \lesssim M_{\rm crit}$, 
which is the same range where the Bondi accretion does not hold (see
Sect.~\ref{sec:valid-bondi-regime}). As it has been discussed there, at these last
stages of the black hole growth, the modifications of the accretion rate can no
longer prevent the star destruction. 

\section{Temperature Considerations}
\label{sec:issue-temperature}

It is clear from the above discussion that because the viscosity depends
strongly on the temperature  $\sim T^{-2}$, the estimates presented above
may not be valid if the accretion onto the black hole increases the
temperature of the surrounding nuclear matter.  Protons, neutrons and
particularly electrons that accelerate towards the black hole radiate due to
thermal Bremsstrahlung or free-free emission. This happens dominantly close to
the event horizon. If a sufficiently large amount of radiation is produced,
two things can happen: a) the heat might increase the temperature of nuclear
matter, making viscosity significantly smaller and therefore invalidating the
mechanism of the evacuation of the angular momentum, and b)  the radiation
pressure may become significant, so that the accretion rate could 
slow down and become much smaller than the
Bondi rate.

The Bremsstrahlung radiation is
produced dominantly close to the event horizon. For Bondi accretion with
$\Gamma=4/3$, the density and temperature at the event horizon are
respectively~\cite{Shapiro:1983du}
\begin{equation}
\frac{n_h}{n_{\infty}} \simeq \frac{ \lambda}{4} 
\left ( \frac{c}{c_s} \right )^3=36, 
\label{nn}
\end{equation}
\begin{equation}
 \frac{T_h}{T_{\infty}}\simeq \left (\frac{\lambda}{4} \right )^{1/3} \left (\frac{c}{\alpha_{\infty}} \right )=3.3, \label{TT}
 \end{equation}
 where we used again $c_s=0.17$ and $\lambda=0.707$. For 
 neutral nuclear matter of  total density $\rho$  in
 $\beta$-equilibrium, one can estimate the Fermi momenta and number densities
 of the neutrons, protons and electrons as follows (see page 310
 of~\cite{Shapiro:1983du})
 \begin{eqnarray}
 n_n= 5.4 \times 10^{38} \left (\frac{\rho}{\rho_0} \right) {\rm cm}^{-3}, \nonumber \\ n_e=n_p=9.7 \times 10^{36} \left (\frac{\rho}{\rho_0} \right )^2 {\rm cm}^{-3}, \nonumber \\ 
 p_{Fn}=0.5 \left (\frac{\rho}{\rho_0} \right )^{1/3} {\rm GeV}, \nonumber \\
 p_{Fe}=p_{Fp}=0.13 \left (\frac{\rho}{\rho_0} \right )^{2/3} {\rm GeV}, \label{nnpp}
 \end{eqnarray}
 where $\rho_0=5 \times 10^{38}~{\rm GeV}/{\rm cm}^3$ is the standard value of
 the neutron star density we have used throughout this paper. 

Let us first notice that as particles flow towards the event horizon, it is
expected that densities will scale by the same factor 36 for all components
(i.e. protons, electrons and neutrons) as given by Eq.~(\ref{nn}). Indeed, the
cross section for weak equilibration is $\sim G_F^2p_FT$ where $G_F$ is the
Fermi constant, $p_F$ the Fermi momentum and $T$ the temperature. The time
scale for weak equilibration is $1/(G_F^2p_FTn)$ where $n$ is the density of
the nucleons that are not Pauli blocked. This time scale is much longer than
the dynamic time scale of Eq.~(\ref{dyn}) and therefore we can safely assume
that  all components scale by the same factor since the weak interactions
are not fast enough to convert species to one another.
 
Let us now estimate the mean free path $d$ of a photon produced close to
the event horizon. It is given by $d=p_{Fe}/(3T_hn_h \sigma_T)$, where
$\sigma_T=6.6 \times 10^{-25}~{\rm cm}^2$ is the Thomson cross section,
$n_h=3.5\times 10^{38}$~cm$^{-3}$ is the electron density at the horizon, and
$3T_h/p_{Feh}$ is the suppression factor due to Pauli blocking of electrons
close to the event horizon. It is understood that $p_{Fe}$ is also evaluated
at the event horizon, which gives $0.43$~GeV. Since for relativistic degenerate matter the number density scales as $p_F^3$,
$p_{Feh}$ is $36^{1/3}$ larger than the asymptotic value of Eq.~(\ref{nnpp}).
 Making use of these numbers and 
Eq.~(\ref{TT}), we get $d =2.2
\times 10^{-8}/T_5$~cm. This has
to be compared to the Bondi radius. If the mean free path $d$ is much
smaller than the Bondi radius, photons are re-absorbed very quickly and no heat
transport from the horizon region to the surrounding nuclear matter takes
place.
A simple comparison shows that this happens for a black hole of mass $M >2
\times10^{43}/T_5$ GeV. Therefore, from this black hole mass to the total
destruction of the star, the Bondi radius is much larger than the mean free
path of photons and no heating of  nuclear matter by the accreted particles
has taken place, beyond the hydrodynamic factor 3.3.

However, this is not sufficient. As explained above, black holes with
masses as low as $\sim 10^{37}$ GeV can be formed out of WIMPs. Moreover,
this case of low-mass black holes is of particular interest as it
corresponds, via Eq.~(\ref{chandra}), to the highest WIMP mass $m$ of
order a few GeV that is considered in many recently proposed WIMP
models.
If the black hole has a mass $M <2 \times10^{43}/T_5$ GeV, photons produced
via thermal Bremsstrahlung of (mostly) electrons can heat the surrounding
nuclear matter and consequently might reduce its viscosity, which may make it
impossible to have a Bondi accretion because the angular momentum is not
effectively subtracted. More importantly, they might slow down the accretion
significantly since the photon pressure can counterbalance the gravitational
attraction from the black hole. 

We consider now this case in more detail. The Bremsstrahlung radiation can be due to
either thermal collisions of protons with electrons, protons with protons and
electrons with electrons. We are going to look at the electron-electron
Bremsstrahlung keeping in mind that the other two combinations are of the 
same order. The Bremsstrahlung radiation in the relativistic case has a
cross section
\begin{equation}
\frac{d\sigma}{d\omega}\simeq \frac{3\alpha \sigma_T }{2\pi E_0^2\omega }
\left ( E_0^2+E^2-\frac{2}{3}EE_0 \right ) 
\left (\ln\frac{2E_0E}{m_e\omega}-\frac{1}{2} \right ), 
\label{brem}
\end{equation} 
where $\alpha=1/137$ is the fine structure constant, $\omega$ is the
energy of the emitted photon, $E_0$ is the initial energy of the
electron and $E=E_0-\omega$ is the final energy. Since the electrons
are degenerate, they must have momentum around the Fermi surface $\sim
p_{Feh}$ where the index $h$ is a remainder that the Fermi momentum
must be evaluated close to the event horizon. For $\omega<<p_{Feh}$,
one has 
\begin{equation}
\frac{d\sigma}{d\omega}\simeq \frac{2 \alpha \sigma_T}{\pi\omega} 
\ln\frac{2p_{Feh}^2}{m_e\omega}. \label{brem2}
\end{equation}
The total emissivity i.e. energy per volume per time is
\begin{equation}
\Lambda_{ee}=\tilde{n}_{eh}^2\int_0^{p_{Feh}} \omega \frac{d\sigma}{d\omega}d\omega, \label{lambdaee}
\end{equation}
where $\tilde{n}_{eh}$ is the electron density close to the event
horizon, including the suppression factor $3T_h /p_{Feh}$ due to the Pauli
blocking since only the electrons that lie within $\sim T_h$ of the
Fermi surface can interact with each other in order to emit a
photon. This is also the reason why the integration stops at
$\omega=p_{Feh}$. Inserting Eq.~(\ref{brem2}) into
Eq.~(\ref{lambdaee}), and performing the integration we get
\begin{equation}
\Lambda_{ee}= \frac{18\alpha}{\pi}\sigma_T n_{eh}^2 
\frac{T_h^2}{p_{Feh}} \ln \frac{2p_{Feh}}{m_e} . 
\label{eee}
\end{equation} 
As before $T_h\simeq 3.3T$, $n_{eh}\simeq 36 n_e$ and
$p_{Feh}=36^{1/3}p_{Fe}$, where the quantities without subscript $h$ refer to the values of Eqs.~(\ref{nnpp})
far away from the horizon. 

The total emitted luminosity is
\begin{equation}
L_{ee}=\int_{r_h}^R\Lambda_{ee}4 \pi r^2dr \sim \Lambda_{ee}\frac{4}{3}\pi r_h^3.
\end{equation}
Because the densities increase as $r$ decreases,
the integral is dominated by the lower bound at $r=r_h$. One can now
estimate the efficiency of radiation, i.e. the ratio of emitted energy
over the total accreted energy, 
\begin{equation} 
\epsilon=\frac{L_{ee}}{dM/dt} \simeq 5 \times 10^{-12}T_5 \left
(\frac{M}{M_0} \right ),
\end{equation}
where $M_0 = 2 \times 10^{43}/T_5$~GeV is the black hole mass above
which the mean free path of the photons becomes smaller than the Bondi
radius and the hydrodynamic approximation holds. Since the infalling
particles are already semi-relativistic and the efficiency is so
small, radiation pressure from Bremsstrahlung cannot impede the
accretion. The out-flowing momentum carried by the radiation is negligible
compared to the in-flowing momentum, and therefore the infalling particles
are not affected by it. 

Consider now the increase of the temperature of the nuclear matter due
to the radiation from the horizon region. A constant
flow of energy $\Lambda_{ee}$ from the center of the star increases
the temperature of the material at a distance $r$
as~\cite{Kouvaris:2012dz}
\begin{equation}
\delta T=\frac{L_{ee}}{4\pi k r}\simeq 458 
\left ( \frac{M}{M_0} \right )^2 \left (\frac{r_B}{r} \right ) {\rm K},
\end{equation}
where $k\simeq 10^3{\rm GeV}^2$ is the thermal conductivity of nuclear
matter~\cite{Kouvaris:2012dz}. Even at the Bondi radius the increase
in the temperature is much smaller than its asymptotic value of
$10^5 - 10^7$~K.  Therefore, we conclude that the
Bremsstrahlung process cannot change significantly the temperature and
does not affect the viscosity mechanism of the angular momentum
evacuation. 

Notice that this result is independent of the asymptotic temperature
for  realistic values of the latter. We should also mention here
that by simple inspection of the relevant equations, Bremsstrahlung
from electron-proton collisions gives identical result whereas the
proton-proton one gives a somewhat smaller contribution (due to the
mass dependence in Eq.~(\ref{eee}) and the fact that
$m_p>m_e$). Therefore, the radiation efficiency is always small and
cannot impede the accretion neither by  radiation pressure nor by
changing the temperature and reducing the viscosity of the surrounding
nuclear matter.

\section{Conclusions}
\label{sec:conclusions}

Stringent constraints have been imposed recently on asymmetric dark matter
candidates based on neutron star destruction by black holes formed out of
collapsing dark matter in the cores of the stars. For these constraints to be
valid, one should make sure that once formed, the black hole consumes the star
in a reasonably short time. This is indeed the case if  spherical Bondi
accretion is assumed. In this paper we have studied the effect of the host
star rotation on the growth of the black hole in its interior. 

If conservation of  angular momentum is assumed for the accreted
matter, we  found that even moderate rotations can disrupt the spherical
accretion at all except the very last stages of the black hole growth. In this case the black hole growth would proceed via a disc-type accretion
with possibly significantly lower rate. This is important for two reasons: 
firstly because the lifetime of the star with the black hole inside might then 
be larger than a few billions years, in which case no constraints can be placed. 
The second reason is that the slow accretion rate may not be enough to win
over  Hawking evaporation at the initial stages of the black hole
growth. Since the accretion rate increases, while the Hawking evaporation rate
decreases with the mass of the black hole, the fate of the latter is determined
by the competition of the two rates right after its birth. A slow accretion
rate means that the black hole would have to be more massive from the start in
order to survive. Thus, knowing exactly the accretion rate at the early
stages of the formation of the black hole is crucial for determining the range
of parameters where the derived constraints are valid.

In this paper we demonstrated that the viscosity of the nuclear matter plays a
crucial role in alleviating the effect of the star rotation, which enables the
black hole to grow via a spherical Bondi-type accretion. More precisely, the
evacuation of the angular momentum due to viscosity is enough to preserve the
Bondi accretion regime from the black hole birth until the moment  it
reaches the mass $M=M_B$ of Eq.~(\ref{eq:mBondi}). Thus, rotation has no effect
neither on the balance between the accretion and the Hawking radiation rates,
nor on the longest initial stage of the black hole growth.

For a large part of the parameter space the growth beyond $M_B$ also proceeds
in the Bondi regime. But for rapidly rotating and hot neutron stars (e.g. $P=5$~ms and
$T=10^7$~K), in the range of masses $M_B<M<M_{\rm crit}$, where $M_{\rm crit}$
is given by Eq.~(\ref{eq:Mcrit}), the Bondi accretion does not hold. However, this does
not prevent the rapid star destruction. The reason is that the Bondi
regime breaks at the very last and short stage of the black hole growth: were
the Bondi regime still valid, the destruction of the star would take less than
a second. We have estimated, by considering  accretion from the polar
regions where the angular momentum is small and is not an obstacle for the
accretion, that the time until destruction cannot be longer than a few hours. The star destruction is thus eminent. 

We also demonstrated that the spin of the black hole initially decreases, but
later on it starts growing and may become close to the maximum one 
 in the mass range $M_B\lesssim M \lesssim M_{\rm crit}$, i.e. when the
spherical Bondi accretion is not valid anyway. In view of the  argument of the previous paragraph,
this has no important consequences for the star destruction time.

Finally, we have investigated whether the Bremsstrahlung radiation of the
infalling matter can impede the accretion or increase the temperature of the
nuclear matter around the black hole in such a way as to modify significantly
its viscosity. We found that the radiation efficiency  (i.e. the
power of the emitted Bremsstrahlung radiation over the accretion rate) is
insignificant and cannot slow down the accretion, while the raise in the
temperature can be at most $\sim 500$~K, which is negligible compared to the
internal temperatures of even  old neutron stars.

\section{Acknowledgements}
C.K. is supported by the Danish National Research Foundation, Grant No. DNRF90.
The work of P.T. is supported by the RFBR grant 13-02-12175-ofi-m 
and Belgian Science Policy under IUAP VII/37.

  \end{document}